\documentclass[rnote]{aa} 

\usepackage{graphicx}
\usepackage{txfonts}
\usepackage{natbib}
\usepackage{longtable}
\usepackage{lscape}

\providecommand{\sorthelp}[1]{}

\begin{document}

\title{Template matching method for the analysis of interstellar cloud structure}

\author{M. Juvela\inst{1,2}}

\institute{
Department of Physics, P.O.Box 64, FI-00014, University of Helsinki,
Finland, {\em mika.juvela@helsinki.fi}                                
\and
Institut UTINAM, CNRS UMR 6213, OSU THETA, Universit\'e de Bourgogne
Franche-Comt\'e, 41 bis avenue de l’Observatoire, 25000 Besan\c{c}on,
France}

\authorrunning{M. Juvela et al.}

\date{Received September 15, 1996; accepted March 16, 1997}

\abstract { 
The structure of interstellar medium can be characterised at large scales in terms of
its global statistics (e.g. power spectra) and at small scales by the properties
of individual cores. Interest has been increasing in structures at
intermediate scales, resulting in a number of methods being developed for the analysis
of filamentary structures.
} 
{
We describe the application of the generic template-matching (TM) method to the
analysis of maps. Our aim is to show that it provides a fast and still relatively robust
way to identify elongated structures or other image features.
}
{
We present the implementation of a TM algorithm for map analysis. The results are
compared against rolling Hough transform (RHT), one of the methods previously used to
identify filamentary structures. We illustrate the method by applying it to $Herschel$
surface brightness data.
}
{
The performance of the TM method is found to be comparable to that of RHT but TM appears
to be more robust regarding the input parameters, for example, those related to the
selected spatial scales. Small modifications of TM enable one to target structures at
different size and intensity levels. In addition to elongated features, we demonstrate
the possibility of using TM to also identify other types of structures.
}
{
The TM method is a viable tool for data quality control, exploratory data analysis,
and even quantitative analysis of structures in image data. 
}


\maketitle

\section{Introduction} \label{sect:intro}

The characterisation of the structure of interstellar medium (ISM) is difficult because
of its complexity. The structures do not generally follow any simple patterns that could
be easily parameterised. The overall properties of interstellar medium can be examined
using general statistics such as the distribution of column density values ($P(D)$
analysis) \citep[e.g.][]{Kainulainen2009, Schneider2015}, power spectra and structure
functions \citep[e.g.][]{Armstrong1981,Green1993,Gautier1992,Padoan2002,Padoan2003}, or
fractal analysis \citep[e.g.][]{Scalo1990, Elmegreen1997}. These methods do not directly
provide information on the shapes of individual objects, only on the global
statistics of a field.

Regarding individual structures, the analysis has concentrated on the smallest scales,
on individual clumps and pre-stellar and protostellar cores
\citep[e.g.][]{Andre1996,Motte1998,WardThompson1999,Schneider2002,Konyves2010}. Because
of the approximate balance of gravity and supporting forces, the shapes at the smallest
scales may be approximated with one-dimensional density profiles or simple 2D shapes
such as 2D Gaussians \citep[e.g.][]{Stutzki1990,Lada1991,Myers1991}, although the
apparent simplicity is sometimes related to the finite resolution of observations.

Recently the interest in the intermediate, better resolved scales has increased. 
The interstellar cloud filaments imaged by the {\it Herschel Space Observatory}
\citep{Pilbratt2010} are one example of this. In the nearby star-forming regions
even the internal structure of filaments can be resolved \citep{Arzoumanian2011,
Juvela2012_filsimu} and detection of filamentary structures is possible up to
kiloparsec distances. In the context of interstellar clouds, the importance of filaments
is partly connected to the role they play in star formation \citep{Andre2014PPVI}.
However, filamentary structures can be detected in almost any kind of astronomical
sources, from the Sun to the distant universe, in both continuum and line data.

Several methods have been used to identify and to characterise elongated structures from
3D data (usually simulations), line observations (position-position-velocity data), and
two-dimensional images. An incomplete list of methods includes DisPerSe
\citep{Sousbie2011, Arzoumanian2011} and getfilaments \citep{Menshchikov2013,
Rivera2016a} routines, derivative-based (Hessian) methods \citep{Molinari2011,
Schisano2014, Salji2015}, curvelets and ridgelets \citep[e.g.][]{Starck2003}, and the
inertia matrix method \citep{Hennebelle2013}. The tools vary regarding their complexity,
computational cost, and the nature and detail of the information they return on the
identified structures.

In this paper, we examine the performance of a simple template matching (TM) technique
in the structural analysis, for detecting for example the presence of elongated
(filamentary) features, in two-dimensional image data. We compare the method to rolling
Hough transform (RHT), which has been used in many image analysis applications
\citep{Illingworth1988} and has been applied to the filament detection from H{\rm I}
observations \citep{Clark2014} and from $Herschel$ continuum data \citep{Malinen2016}.
RHT was selected because its basic principles are very similar to the proposed TM
implementation. Both methods are used to enhance certain structural features of the
input images and thus to help characterise the structures, for example,
regarding the degree of angular anisotropy. Neither method is geared towards the
extraction of individual sources.

Our aim is to show that TM provides a fast and yet relatively
robust way to identify local anisotropy and that, with minor changes, can also be
used to highlight other types of structures. The content of this paper is as
follows. Section~\ref{sect:methods} gives an overview of the TM and RHT methods.
Further details on our TM implementation are given in
Appendices~\ref{appendix:implementation} and \ref{appendix:alternative}. In
Sect.~\ref{sect:results} we present a series of tests where the results of RHT and our
TM implementation are compared for synthetic images. We discuss the results and possible
applications of the TM method in Sect.~\ref{sect:discussion} before summarising our
conclusions in Sect.~\ref{sect:conclusions}.
\footnote{Our python implementation of the TM routine can be obtained from {\tt
www.interstellarmedium.org/PatternMatching}.}

\section{Methods} \label{sect:methods}

\subsection{Template matching} \label{sect:TM}

Our aim is to highlight features that are similar to a given generic pattern in the
input images. Pattern matching (or pattern recognition) is a broad category of methods
to identify regular features of input data \citep[see, e.g.][]{Manfaat1996}. In
particular, we follow the template matching (TM) approach \citep{Brunelli2009}
where data are compared to only a single pattern that is also kept fixed (apart from
rotation). Such templates are limited to primitive forms (e.g. gradients or elongated
shapes) contrary to many computer vision applications (see, for examples, the algorithms
in the OpenCV package\footnote{http://opencv.org}). The data are not expected to contain
exact copies of the template. Thus, each match is associated with a significance number
and the process can also be seen as a form of regression \citep{Bishop2006}.

First, we look for a way to emphasise elongated structures of a given size
scale in images. We start by creating two images that are obtained by convolving the input
image with Gaussian beams with full width at half maximum (FWHM) values of one and
(by default) two times a given size parameter $F_1$. The lower resolution image is
then subtracted from the higher resolution one. In the second stage, we seek
structures in this difference image, that contain a limited range
of spatial scales. We compare at each pixel position the data with a pre-defined
template. A simple template applicable to the detection of elongated structures of
excess signal (and the one that gives zero output for a flat input image) is
\begin{equation}
\begin{pmatrix}
 -1/3  &  2/3  &  -1/3 \\
 -1/3  &  2/3  &  -1/3 \\
 -1/3  &  2/3  &  -1/3 \\
\end{pmatrix}
\label{eq:template}
\end{equation}

The template or stencil is placed on the map and the data values for the
corresponding nine positions are read from the difference image. The step size between
the stencil elements is taken to be equal to the value of the $F_1$ parameter. Thus,
$F_1$ determines the angular scales for which the method is most sensitive. The data
values are multiplied by the corresponding template elements and the sum is
interpreted as the significance $S$ of the match between the template and the data.

The template is rotated over a range of possible orientations. If the template is not
symmetric, the orientation can cover 360$\degr$ and for symmetric templates, as
in Eq.~(\ref{eq:template}), the angle is between $-90\degr$ and $+90\degr$. For the
above template, zero position angle would correspond to a vertically elongated structure
(we measure position angles counter-clockwise from north). In the tests, we
sample angles with a step of 3$\degr$ or less, which corresponds to a finer sampling
than that suggested by Eq.~(2) in \citet{Clark2014}.

The procedure is repeated for each pixel position. Although the
significance $S$ could be saved as a function of the position angle or calculated
as a weighted average over different angles (in a way similar to RHT, see below), we
simply compress the result of the analysis into two maps, one for the position angle
$\theta$ of the best match (highest $S$ value) and one for the corresponding
significance $S$. In our calculations, the resulting maps have the same dimensions
as the input maps although the pixel size could be larger, in
proportion to the $F_1$ value.

Direct multiplication of an input map with the template gives larger weight for pixels
with larger absolute values. Thus, the $S$ values depend on the average signal level and
can be sensitive to outliers. We can alternatively normalise the data under the current
template by dividing these by the standard deviation of this sample of nine data
points (for the current position and orientation of the template). In the normalised
version (TMN) the $S$ values depend mainly on the shapes of the template and input data,
less on their absolute values.

As in the case of RHT (see Sect.~\ref{sect:RHT}), the main result is the combination of
the significance and position angle maps. The significance values can be used in
two ways. First, one can select for plotting a certain percentage of pixels with the
highest $S$ values. It is necessary to mask most noisy pixels so that the image can be
inspected visually for the presence of regions with high $S$ values and coherent
position angles. This is sufficient for general data exploration and works best when,
as in the case of Eq.~(\ref{eq:template}), 
the matched structure can be long compared to the size of the individual template, thus
creating large continuous regions of high significance.
Second, one can make quantitative estimates of the
statistical significance of a given $S$ value by using the expected distribution of $S$
values. If the data only contain white noise and background fluctuations with a standard
deviation of $\sigma_0$ at the scale $F_1$, the template of Eq.~(\ref{eq:template}) will
result in a distribution $N(0,\sqrt{2}\sigma_0)$ of $S$ for a random position angle.
Here $N(0,\sqrt{2}\sigma_0)$ refers to a normal distribution with a mean of zero and a
standard deviation of $\sqrt{2}\sigma_0$, which results from the sum over the nine
template elements in Eq.~(\ref{eq:template}). For TMN the corresponding distribution is
$N(0,\sqrt{2})$ (except that the normalisation is carried out using
$\sigma_0$ estimated from only nine data points). 

The actual relationship between $S$ and statistical significance is modified by two
additional factors. First, the template is matched to filtered images, that decreases
the $S$ values (for $F_1$ larger than one pixel). Second, at each location TM selects
the position angle that gives the largest $S$ value. This makes the expectation value of
$S$ positive and skews the probability distribution. Thus, the real significance depends
in a non-trivial way on the spectrum of background fluctuations (as a function of
spatial frequency), the symmetry properties of the template, and the number of tested
position angles. The mapping between $S$ and probability can be determined with a Monte
Carlo simulation. For example, for pure white noise with $\sigma_{\rm pix}=1$ per pixel
and with parameters $F_1=2$ and $F_2=4$ pixels, $S$ values of 1.11, 1.54,
and 2.03 correspond, respectively, to 1$\sigma$, 2$\sigma$, and 3$\sigma$ detections
(confidence limits of 84\%, 97.7\%, and 99.87\%). A given percentile of $S$ values from
the analysed image corresponds to a confidence limit that is equal or higher, because
the presence of real structures adds a tail of high values to the $S$ distribution.

The combined significance of a continuous region with coherent position angles is 
higher. The probabilities associated to $S$ values can be combined in the usual manner,
taking into account that estimates are independent only at distances larger than the
template size. The significance is further increased by the evidence of the similarity
of position angles. A random input should result in a uniform distribution of $\theta$
values, except for potential pixel discretisation effects when $F_1$ approaches the
pixel size. The selection of the tested region itself can also bias the estimates of the
combined significance. We return to the question of combined probabilities briefly in
Sect.~\ref{sect:reliability}.

As a next step of the analysis, one can trace the skeletons of filamentary
structures using the $S$ and $\theta$ data. This applies only to the particular template
of Eq.~(\ref{eq:template}) and and we do not consider this to be a part of the TM method
itself. In this paper we use skeletons only for illustration purposes and to assist in
testing of the method (see Sect.~\ref{sect:reliability}). Our skeleton tracing algorithm
has three parameters, $S_{\rm H}$, $S_{\rm L}$, and $\Delta \theta_{\rm max}$. The
pixels associated with a skeleton must all have $S>S_{\rm L}$ and there must be at least
one pixel with $S>S_{\rm H}$. Furthermore, $\Delta \theta_{\rm max}$ specifies the
maximum allowed change of the position angle between consecutive skeleton positions. The
ridge tracing proceeds in steps of $F_1$, always finding the maximum $S$ value within
the allowed $\Delta \theta_{\rm max}$ sector around the local estimate of the position
angle.

\begin{figure}
\includegraphics[width=8.8cm]{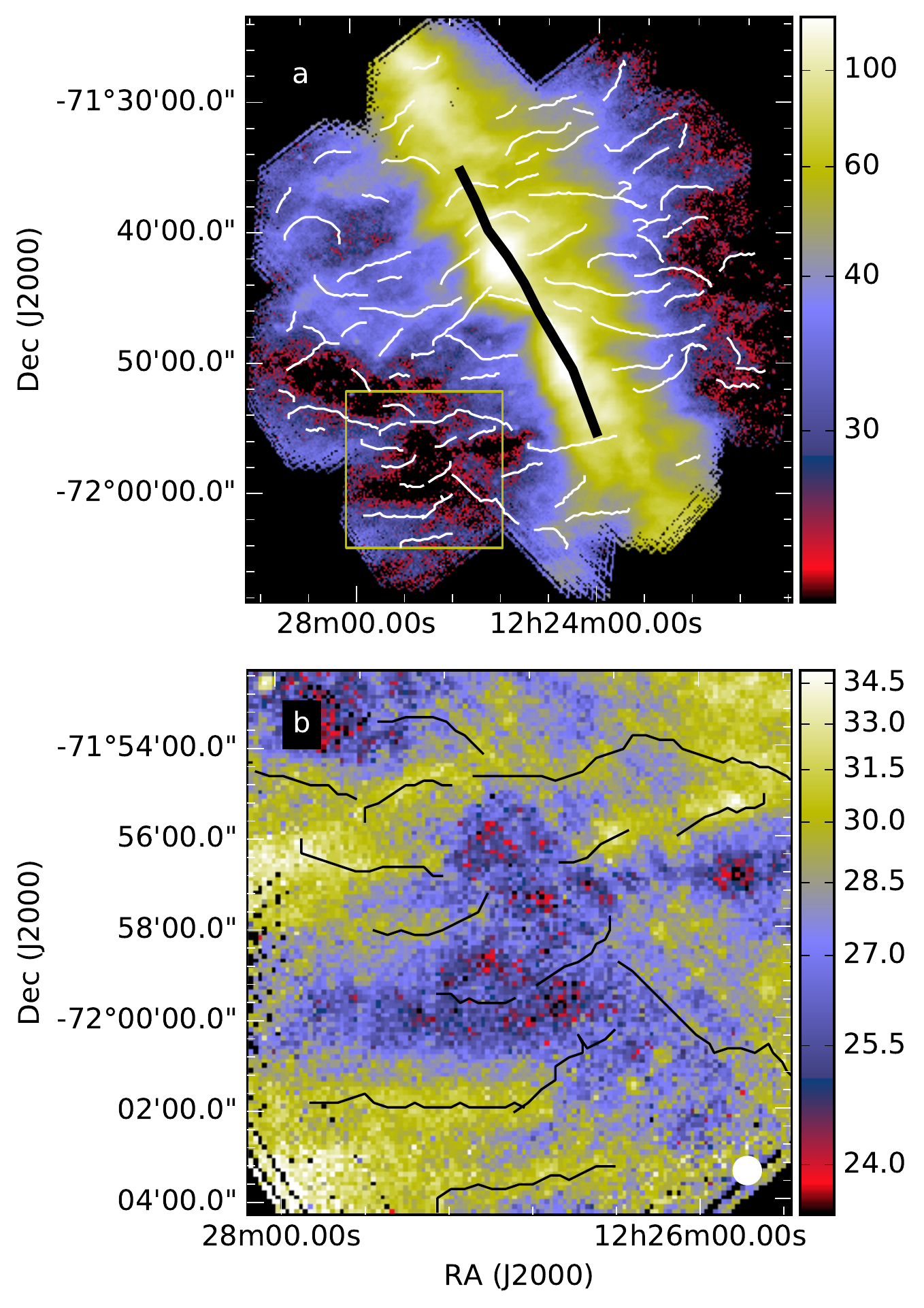}
\caption{
Filament skeletons based on TM results. The skeletons correspond to a scale of
$F_1=0.6\arcmin$ and the TMN method except for the thick black line that shows the only
skeleton found at the scale of $F_1=5.0\arcmin$ and without data normalisation. The
background is the $Herschel$ $250\mu$m surface brightness image in units of
MJy\,sr$^{-1}$. The lower frame shows details of a smaller region (marked with a box in
the upper frame) where the $F_1=0.6\arcmin$ skeletons are drawn in black for better
contrast.
}
\label{fig:skeleton}
\end{figure}

Figure~\ref{fig:skeleton} uses skeletons to illustrate the qualitative effect of the
selected scale and normalisation. The data are the $Herschel$ 250\,$\mu$m surface
brightness observations of the field G300.86-9.00 in Musca. The map is the one used in
papers \citet{GCC-V, GCC-VI}. At large scales (and high column densities) the field
shows a single filament. On the other hand, at small scales and lower column densities,
one can also recognise faint striations that have a clearly different distribution
of position angles. In the case of TMN, one must be careful that the results for faint
structures are not affected by artefacts like striping. In Fig.~\ref{fig:skeleton}, only
few skeletons are aligned with the scan directions. The skeletons are drawn using
$S_{\rm H}$ and $S_{\rm L}$ values that correspond to the 96\% and 80\% percentiles of
the $S$ distribution, requiring a minimum length of $3 \times F_1$. With these settings,
even the faintest extracted structures seem to correspond to real features in the data
(Fig.~\ref{fig:skeleton}$b$).

\subsection{Rolling Hough transform} \label{sect:RHT}

We compare our TM results to the results of the RHT analysis. The RHT method
is described in detail in \citet{Clark2014} so we include here only a brief overview.

The RHT procedure uses the difference of the original map and a lower resolution map
that is obtained by convolving the data with a top-hat kernel with diameter $D_{\rm K}$.
One examines a circular region with a diameter $D_{\rm W}$ that is centred on position
$(x,y)$. One counts the number of positive pixels, $R(\theta, x, y)$, that fall under a
straight line that crosses the centre point of the region and has a position angle
$\theta$. Using a pre-defined parameter $Z$, the directions with the number of positive
pixels below $Z \times D_{\rm W}$ are ignored. The significance of the detection of an
elongated structure is described by an integral of $R(\theta, x, y)$ over
360$\degr$ and the mean position angle is obtained as
\begin{equation}
<\theta> = \frac{1}{2} \arctan 
\left(
\frac{\int \sin(2\theta) R(\theta) d\theta}{\int \cos(2\theta) R(\theta) d\theta}
\right).
\label{eq:RHT_theta}
\end{equation}
This calculation of the mean position angle can be done at each pixel
position or as an average over a large region. For further details, see
\citet{Clark2014}.

\section{Results} \label{sect:results}

\subsection{Comparison to RHT} \label{sect:results_RHT}

In this section, we apply our TM method to simulated images. The results are compared
with the results of the RHT method. The images are 360$\times$360 pixels in size and the
methods are set to identify elongated structures that are at least ten pixels long. Thus,
in the TM method the length of the template ($2\times F$) is 10 pixels and in RHT
the parameter $D_{\rm W}$ is similarly 10 pixels.

\subsubsection{Elongated clumps with white noise}

In the first test, we simulate a series of elongated clumps on a flat background and with
white noise with standard deviation equal to 5\% of the peak intensity of the
structures. The intensity distribution of each clump follows a two-dimensional Plummer
type profile with an additional exponential cut-off that results in a more 
square profile
along the clump main axis. The intensities are calculated from the formula
\begin{equation}
I(x,y) = \left[ 1 + (x/R_1)^2 + (y/R_2)^2 \right]^{-p} \exp(-0.05(x-R_1)^2).
\label{eq1}
\end{equation}
The parameters $R_1$ and $R_2$ determine the length and the width of a clump,  the
coordinates ($x$,$y$) having their origin at the centre of the clump.

\begin{figure*}
\includegraphics[width=18cm]{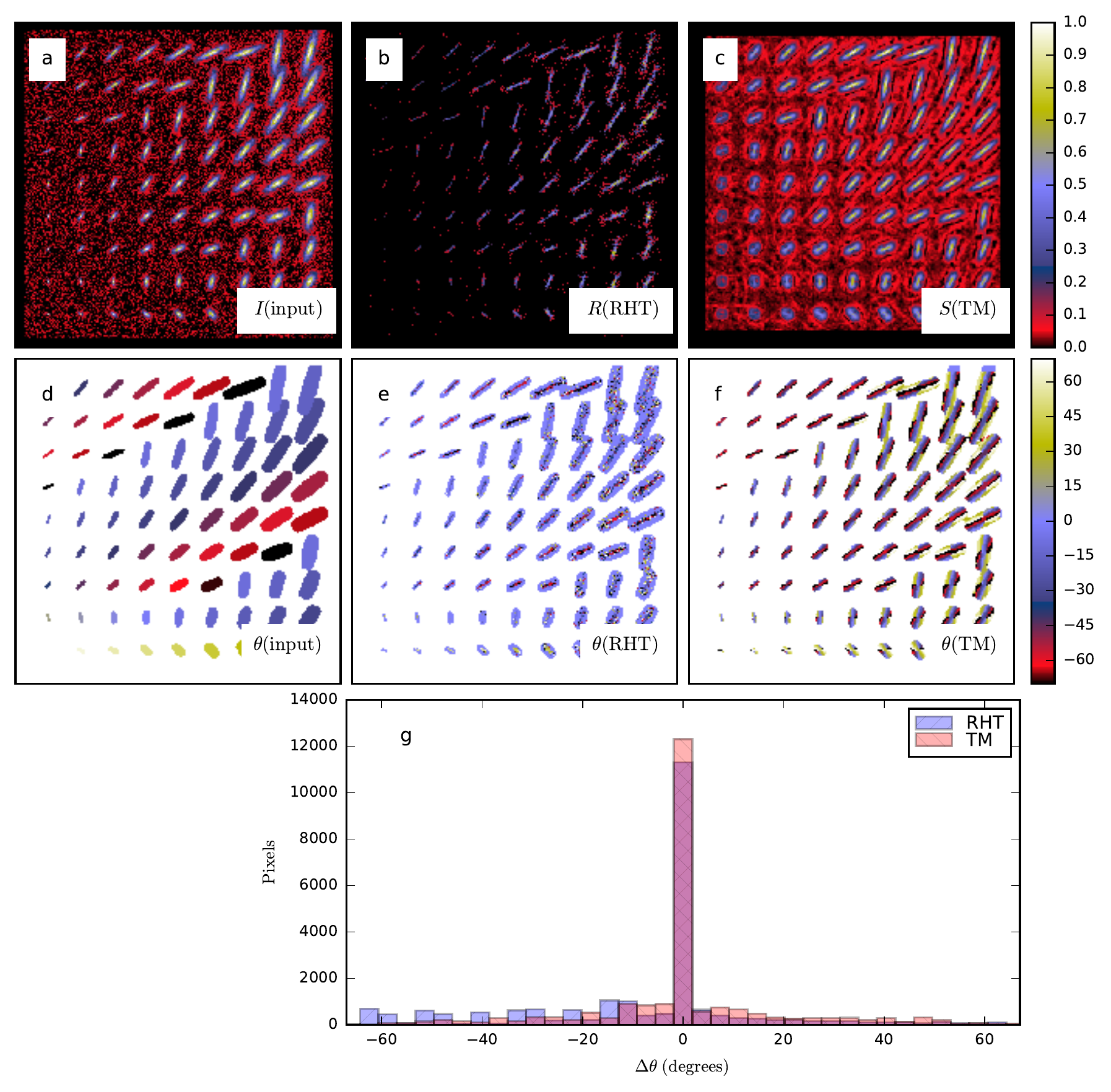}
\caption{
Comparison of RHT and TM in case of short elongated structures. 
Frame $a$ shows the
input image and frame $d$ the position angles (in degrees) of the clump main axes.
The width parameter $R_2$ increases from one to five across the columns of clumps. The ratio
of $R_1/R_2$ (elongation) is varied between the values of two and five, the uppermost row of
clumps having the largest elongation.
The
second and third column correspond to RHT and TM results, respectively. The RHT map of
$R$ is shown in frame $b$ and the TM map of $S$ in frame $c$. The recovered position
angles are shown in frames $e$ and $f$, for RHT and TM, respectively. In the second row
maps pixels with the signal below the 1-$\sigma$ noise level have been masked. The frame
$c$ shows histograms of position angle errors, in correspondence to the data in frames
$e$ and $f$.
}
\label{fig:simu_1}
\end{figure*}

The first frame in Fig.~\ref{fig:simu_1} shows the input map that contains a grid of 81
clumps of different sizes and position angles. All clumps have the same peak intensity
and the value of $p=2$ but the parameters $R_1$ and $R_2$ are varied. The maximum length
of the structures is about 30 pixels and the full width (measured as signal above the
noise level) is at most approximately ten pixels, equal to the size of the regions analysed by
both RHT and TM. 

The second and the third columns of Fig.~\ref{fig:simu_1} show the significance (scaled
between zero and one) and position angle maps. All quantities are calculated pixel by
pixel, without any spatial averaging.  Exact position angles are usually recovered only
close to the clump centre line.
For RHT, the angles tend to zero whenever the signal is low, in weak clumps
and in the outer parts of all clumps. For TM, the angle changes by 90$\degr$ at the
outer clump borders. The difference between the methods is caused by the fact that we
adopt for TM the single position angle that resulted in the largest $S$ value. As shown
in Fig.~\ref{fig:simu_1}, the histograms of position angle errors are very similar
between the two methods and, in particular, there is about the same number of pixels for
which the correct position (within $\sim$2$\degr$) was recovered. 

\subsubsection{Elongated clumps on fluctuating background}

\begin{figure*}
\includegraphics[width=18cm]{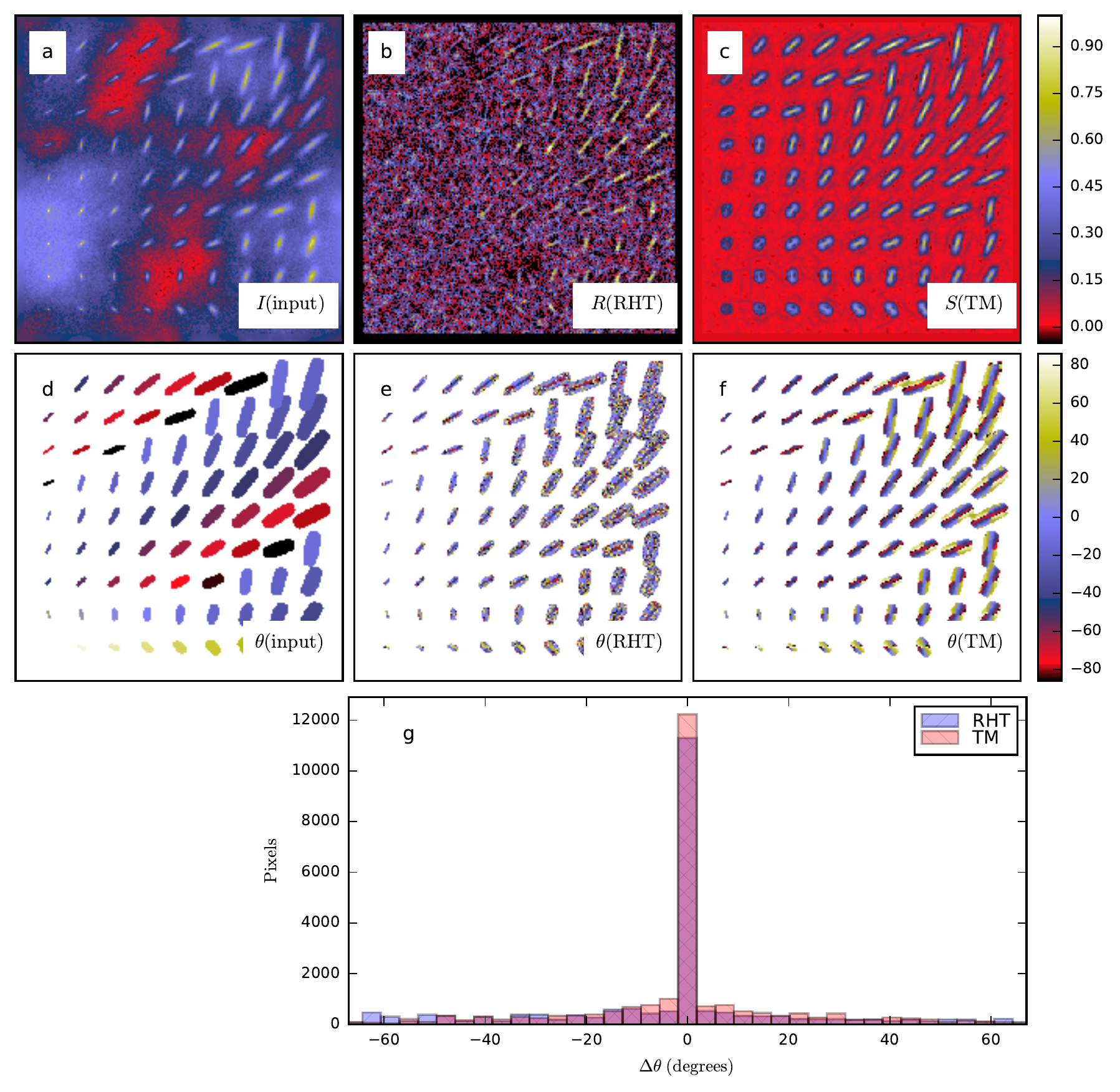}
\caption{
Comparison of RHT and TM results in the case of a fluctuating background.  For the
description of individual frames see Fig.~\ref{fig:simu_1}.
}
\label{fig:simu_2}
\end{figure*}

In the second test we add not only white noise (again with rms noise equal to 5\%
of the peak intensity of the clumps) but also background fluctuations that follow a
power law $k^{-2}$ as a function of the spatial frequency $k$ (spatial scales from one
to 180 pixels). The results are shown in Fig.~\ref{fig:simu_2}.
We keep the parameter values of both methods the same as in the previous test of
Fig.~\ref{fig:simu_1}. For TM the values are $F_1$=5 pixels and $F_2/F_1=2$ and
the parameters of RHT are $D_{\rm K}$=10 pixels, $D_{\rm W}$=5 pixels, and
$Z$=0.6. We also checked that a change of any single RHT parameter did not lead to any
significant improvement of the result.

Compared to the RHT $R$ quantity, the TM map of $S$ (Fig.~\ref{fig:simu_1}c) shows more
clearly the locations of the smallest clumps. This is partly because the input data are
not thresholded, the actual values of individual pixels contribute to the $S$ values,
and the signal-to-noise ratio is still relatively high. In TM the detection was done on
the difference of two maps whose spatial resolutions differing only by a factor of two.
Without the high pass filtering the large-scale gradients increase the dispersion of the
$S$ estimates. However, in practice the effect is negligible for pixels with 10\% of the
highest $S$ values.  TM is able to determine correct position angles within 2$\degr$ for
more pixels than RHT does. There are also more pixels, for which the angles are
within $\pm$20\%. Even these estimates are still useful because the error is well below
the standard deviation of $52\degr$, which applies to completely random angles
that are limited to within 90$\degr$ of a given direction.

\subsubsection{Filaments on fluctuating background}

In the final test with synthetic maps, we use simulations of long nonlinear
filaments. Filaments are in this context useful as visual entities, but for
both methods the calculation is carried out independently at every pixel position. The
cross-filament profiles are generated using a Plummer function with exponent $p=2$, 
similar to Eq.~(\ref{eq1}) but as function of $y$ only. The width parameters $R_2$ are
generated as random numbers $3 \times (U(0,1)+0.5)$ (in pixel units), where $U(0,1)$ refers
to uniform random numbers between zero and one. This gives a vector of $R_2$ parameter values
that correspond to equidistant locations along the filament.
The values are smoothed (as a function of the filament length) using a 1D Gaussian
filter with FWHM equal to three pixels. The ridge values of the filaments are generated in a
similar fashion, using random numbers $N(0,1)^2$ (squared simply to obtain a
non-negative number) to which we apply 1D Gaussian convolution (as a function of the
position along the filament) with FWHM equal to 3 pixels. $N(0,1)$ are normal
distributed random numbers with a mean of zero and a standard deviation of one. The
filament orientation is generated as a random process that allows rapid changes of
direction. 
The initial position of the filament is taken from a uniform 2D distribution of
positions over the map area and the initial direction of the filament is selected as a
random number of $2\pi U(0,1)$ radians. We generate a vector of 240 random numbers
$\theta=0.2\times N(0,1)$ and carry out 1D Gaussian convolution of this vector using a
filter with a standard deviation $\sigma=$4. The elements of the $\theta$ vector are
then interpreted as directions of 240 consecutive steps along the filament, each with a
length of 0.5 pixels. Finally, each filament is shifted so that it is contained within
the map area.

\begin{figure*}
\sidecaption
\includegraphics[width=18cm]{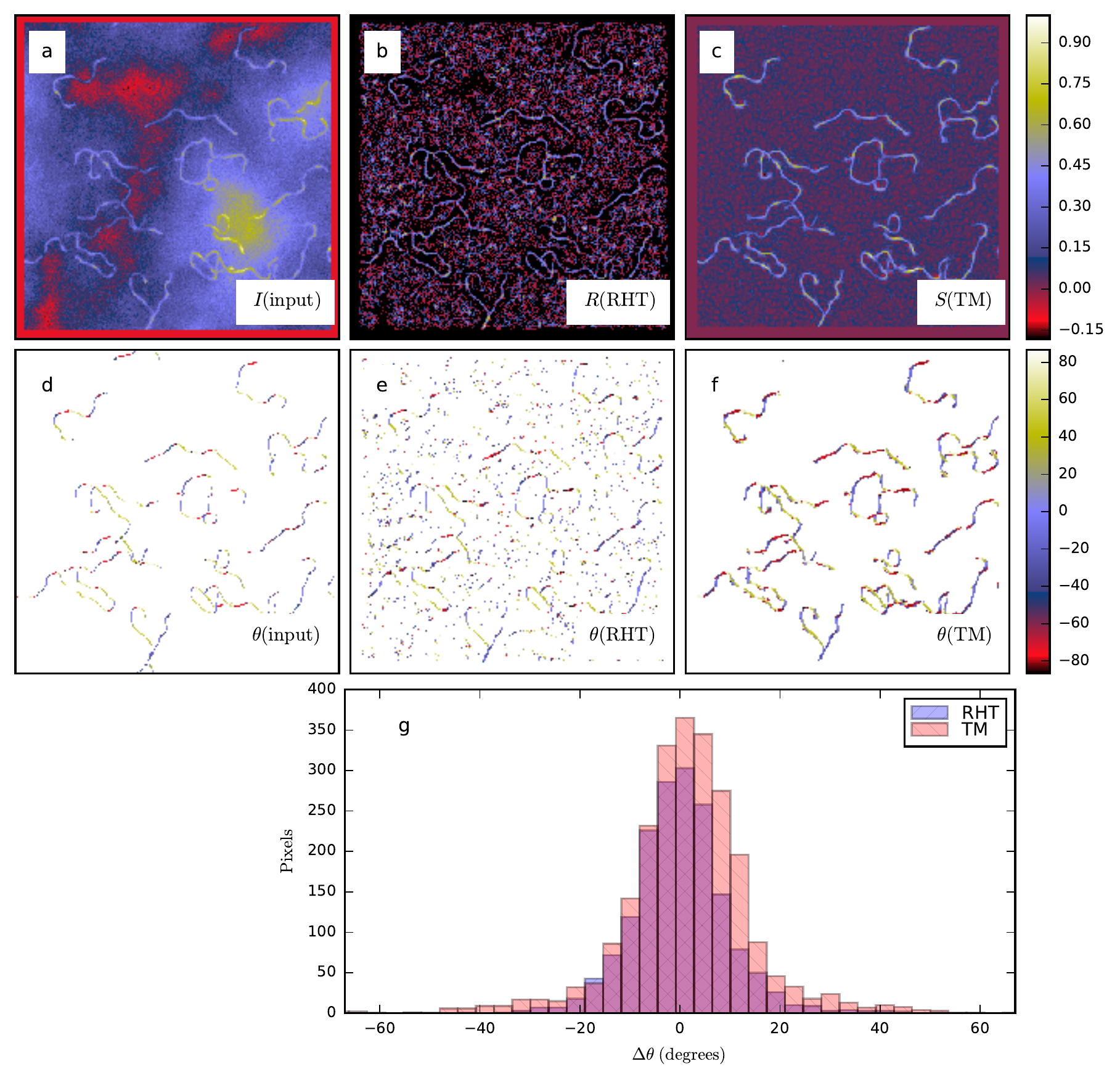}
\caption{
Comparison of RHT and TM results in the case of nonlinear filamentary structures on a
fluctuating background. For the description of individual frames see 
Fig.~\ref{fig:simu_1}.
}
\label{fig:simu_3}
\end{figure*}

Figure~\ref{fig:simu_3} shows one simulation. The noise level is higher than in the
previous examples and its standard deviation is equal to 16\% of the expectation value
of the mean intensity along the filament ridges. Therefore, the significance maps are
less reliable and not all pixels above even a high percentile point 
necessarily belong to a real filament. The histograms in Fig.~\ref{fig:simu_3}g are
plotted for the intersection of two sets of pixels, those that are above the 95\%
percentile of significance (given by the $R$ and $S$ values) and those that coincide
with a real filament (see Fig.~\ref{fig:simu_3}d). The RHT parameters are this time
optimised manually, only keeping $D_{\rm W}$=10 fixed. This leads to values $D_{\rm
K}$=5 and $Z$=0.65. By visual inspection of the results, TM is slightly more successful
in highlighting the location of filaments (Fig.~\ref{fig:simu_3}c) and estimating their
position angles (Fig.~\ref{fig:simu_3}g).

\begin{figure*}
\includegraphics[width=18cm]{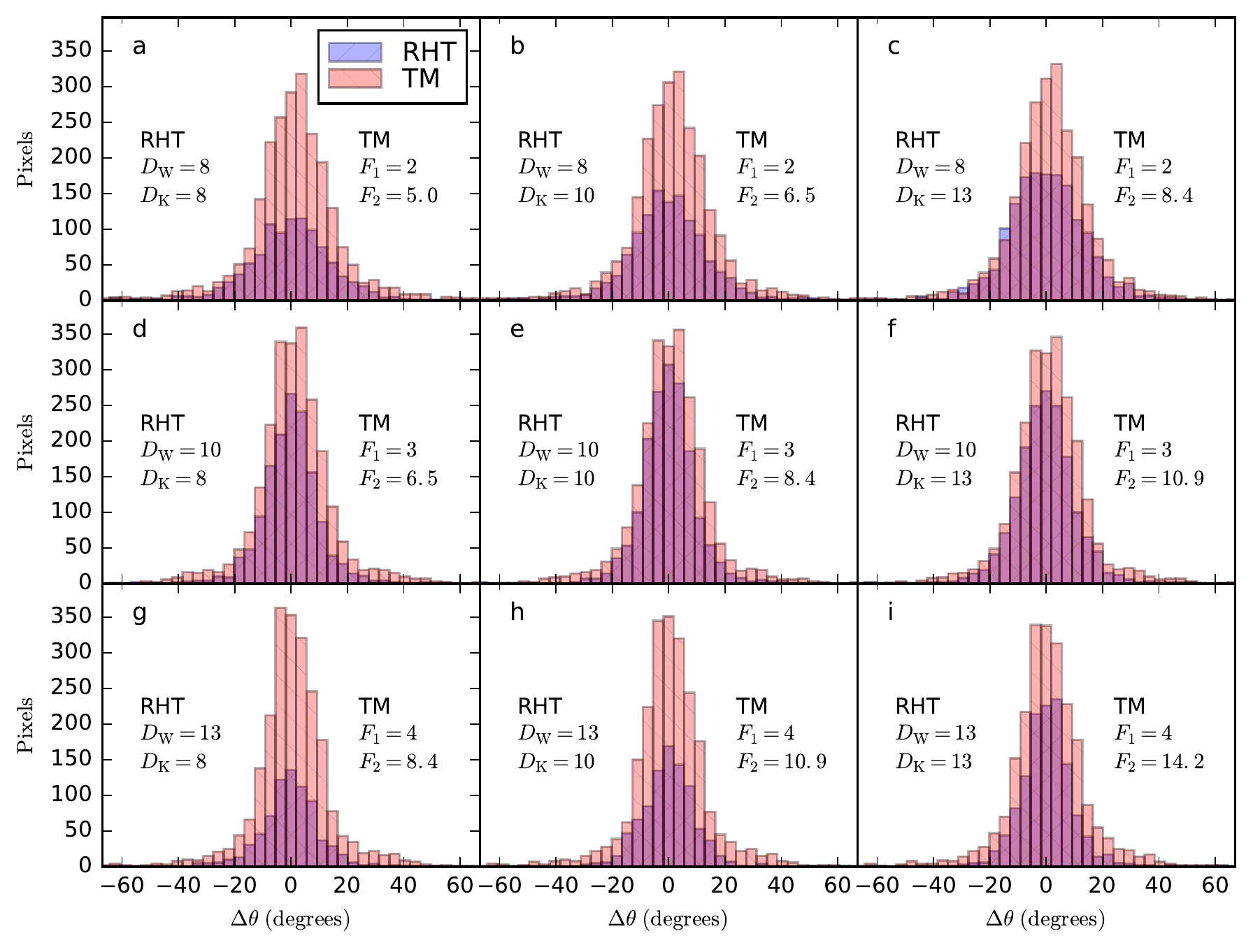}
\caption{
Change in RHT and TM results as a function of the method parameters. RHT parameters
$D_{\rm W}$ and $D_{\rm K}$ are varied and parameter $Z$ is kept at a value of 0.65. 
}
\label{fig:simu_3_grid}
\end{figure*}

Figure~\ref{fig:simu_3_grid} shows how the results change as a function of the parameter
values. For TM, the values of $F_1$ and $F_2/F_1$ are varied in 30\% steps. For RHT, the
parameter $Z$ is kept at a value of 0.65, which was found to give the best results in
Fig.~\ref{fig:simu_3}, but $D_{\rm W}$ and $D_{\rm K}$ are also change in 30\% steps
around the best parameter combination. The results of TM are less
sensitive to these changes in the input parameters.

\subsection{Reliability and completeness} \label{sect:reliability}

We address questions of TM reliability estimation in the case of the simulation shown in
Fig.~\ref{fig:simu_3}. $F_1$ equals 2.8 pixels and any two estimates that are separated
by approximately five pixels can be considered independent. We first analyse maps
similar to those in Fig.~\ref{fig:simu_3} but without filaments and with different
realisations of noise and of large-scale fluctuations. According to this Monte Carlo
exercise, 84\%, 97.7\%, and 99.9\% confidence limits correspond to $S$ values of 0.0457,
0.0638, and 0.0850, respectively. This can be compared to the $S$ values of the same
percentiles that are obtained directly from the input map of Fig.~\ref{fig:simu_3}.
Those values are 0.057, 0.23, and 0.43 and they would thus result in lower estimates of
the statistical significance of a given $S$ value. 
For example, $S=0.057$, the apparent value corresponding to the 84\% confidence limit,
is in reality closer to the 97.7\% confidence limit. The $S$ values are higher because
of the presence of real elongated structures. The more real structures the map has, the
more conservative are the $S$ limits derived from the statistics of the image itself.

For a 360$\times$360 pixel image, one would expect 2981 spurious pixels above the given
threshold of $S_0=0.0638$. We count as spurious all pixel whose distance to real
filaments exceeds three pixels. The number of such pixels is 3174, by 6.5\% higher than
expected. However, there are still some pixels with $S>S_0$ that are associated to
filaments but are at distances exceeding three pixels. This suggests that the $S_0$
threshold obtained from the Monte Carlo simulations gives a rather accurate estimate of
the reliability. For the 97.7\% confidence limit, we expect the map to contain 68
instances where two spurious detections are at a distance of $F_1$ of each other.
Additionally, their position angles would also be expected to match, for example, by
$2\degr$ only once per map. Thus, even the 97.7\% confidence limit should in practice
lead to a high probability for any structure that extends over several template lengths.

The second point we examine is the completeness of detections along the filament ridges
and the rms error of the corresponding position angle estimates. For this purpose, we
use filament skeletons extracted using parameters $S_{\rm H}=0.0850$ and $S_{\rm
L}=0.0638$, and $\Delta \theta_{\rm max}=45\degr$. The completeness, defined as the
fraction of the length of true filaments matched by a recovered skeleton within $\pm$3
pixel distance, is 95.2\%. We do not show a figure of the skeletons, which closely
match the filaments that are seen in the input map in Fig.~\ref{fig:simu_3}. The rms
error of the position angle estimates along the recovered skeleton is 17.5$\degr$. The
number includes the larger errors from the locations of filament crossings. It could
also be affected by imperfections in the filament tracking but the rms error is not
lower if calculated for pixels along the true filament ridge. It is possible to reduce
the position angle errors further by using the information that is contained in the $S$
map and that is also carried on to the skeletons. However, in our case we trace the
skeleton only at steps of $F_1$ and at an accuracy of one pixel. This is not sufficient
for accurate estimates of the local tangential direction, especially in the present case
where the filaments exhibit rapid changes of direction.

\section{Discussion}  \label{sect:discussion}

The template matching (TM) approach was found to provide a simple way to locate and to
highlight pre-defined features in image data. A comparison with RHT demonstrated
that the TM method is efficient in identifying elongated structures and in estimating
their orientation. This is at least partly a result of the decision not to threshold the
data that are used to calculate the significance values. The tested maps (both simulated
and real observations) had a relatively high signal-to-noise ratio and data quantisation
(via thresholding, as in the case of RHT) would result in loss of information.

We carried out tests with and without normalisation of data with its standard deviation.
Without normalisation, the results are closer to RHT and the extraction concentrates on
the highest intensity structures.  In $Herschel$ maps, a surprising amount of detail is
revealed even in diffuse regions when data normalisation is used. This is the case for
example in Fig.~\ref{fig:alt_G210_90-36_55_std1_filter1} that corresponds to the
lowest column densities $N({\rm H}_2)$ of only a few times 10$^{20}$\,cm$^{-2}$. The
coherence of the position angles, which usually extends over a distance of several
times the size of the template, strongly suggests that a large fraction of all plotted
image pixels belong to real structures. Thus, TM can be used not only to highlight
individual structures but also to demonstrate general anisotropy of the surface
brightness or column density distribution.

We concentrated on the TM results computed for each pixel position separately. The
visual inspection of the results was based on $S$ and $\theta$ maps thresholded at a
given $S$ level. A low $S$ limit means that images also contain spurious individual
pixels (pixels that are above the threshold but are not related to any significant
structure). However, the human eye is good in recognising coherence, that is, separating
potentially significant features from the noise. Images may also show general anisotropy
that is visible in the presence of numerous faint features, none of which are
significant individually. This can be sufficient for a qualitative characterisation of
images when TM is applied simply as a pre-filter. However, as discussed in
Sect.~\ref{sect:TM}, it is also possible to derive quantitative estimates of the
statistical significance of individual $S$ values.

In the case of the filament template, the skeletons of the filaments can be traced
based on the information contained in the $S$ and $\theta$ maps (e.g, in
Fig.~\ref{fig:skeleton}) but this involves additional choices (for example, minimum $S$
values accepted in a filament or maximum allowed changes of $\theta$) beyond the
pure TM method. The results of TM already point to the presence of overlapping filaments
(see, for example, Fig.\ref{fig:alt_G210_90-36_55_std1_filter1}) that should be common
for any long line of sight through the interstellar medium. Ideally, filament
tracing would be able to handle also the case of crossing filaments, retaining their
identity as separate objects \citep[see][]{Breuer2016}. In practice, the distinction
between a sharp turn of a filament and the crossing of two filaments is not easy to
make. Once a structure (e.g. a skeleton) has been extracted, its reliability as a
separate object can be evaluated based on the $S$ values of the individual pixels and
(depending on the template) the continuity of the position angle estimates (see
Sect.~\ref{sect:reliability}).

We have presented results for selected scales that in the case of $Herschel$ maps ranged
from 18$\arcsec$ to 5$\arcmin$. However, TM could be run routinely for the full range of
scales from data resolution to the scale of the entire image. At each resolution, one
could also make use of all available data that may exist for different frequency
bands. The $S$ and $\theta$ maps of individual images could be combined, for
example, by calculating averaged $\theta$ values weighted by $S$ and by calculating a
combined $S$ in a way similar to the way RHT combines $R$ values of different $\theta$.
If all maps are assumed to contain evidence of similar structures, the analysis could of
course be carried out simply for a weighted average of the input maps. In case of
$Herschel$, at scales above $40\arcsec$, one possibility is to use column density maps
that are already a combination of data at several wavelengths. This would be a natural
option, if the analysis targets the cloud mass distribution.

We used simulated and $Herschel$ surface brightness maps with map sizes up to about
$900\times 900$ pixels. The run times of a single analysis were of the order of one
second, even when the match was (somewhat excessively) repeated in an oversampled manner
at each pixel position.  The computational efficiency makes TM suitable for
exploratory data analysis, as the first step that can then potentially point out the
need for further analysis with other methods. Via position angle histograms, it is also
a viable tool for the general characterisation of the anisotropy of the structures
within an image. Other applications could be found in semi-automatic quality control of
images where templates could be constructed to specifically detect certain kinds of
artefacts. The row/column artefacts of CCD images could be one example, where a match
would also need to be tested only for two position angles.

TM is a generic framework and an analysis could involve not only different scales
but also combinations of the results with different templates. With minor modifications,
the basic algorithm is applicable also in other dimensions, in particular for 1D and 3D
data. In a future paper, we will use the TM method to characterise the general
structural anisotropy of the $Herschel$ images observed in the Galactic Cold Cores
project.

\section{Conclusions}  \label{sect:conclusions}

We have studied the use of a simple template matching (TM) algorithm to highlight
structural elements in two-dimensional images. We examined the sensitivity of the method
to detect elongated structures, for example, in $Herschel$ surface brightness
data. In case of synthetic maps, the results were compared to those obtained with
RHT. The study leads to the following conclusions:
\begin{itemize}
\item Template matching techniques provide a simple and yet an efficient method of image
analysis in terms of both computational cost and the quality of results.
\item In detection of elongated structures, TM provides results comparable to
those of RHT. However, TM was found to perform better when maps contain significant
noise and background fluctuations.
\item Small changes of the basic algorithm (e.g. regarding data normalisation and spatial
filtering) can be used to emphasise the brightest structures or to highlight
structures at very faint levels.
\item In the case of $Herschel$, its data quality allows the identification of
filamentary structures at very low surface brightness levels, down to column
densities $N({\rm H}_2) \sim 10^{20}$\,cm$^{-2}$.
\end{itemize}

\begin{acknowledgements}
MJ acknowledges the support of the Academy of Finland Grant No. 285769. 
We thank Tuuli Kirkinen for useful discussions and help in the preparation of the
manuscript.
\end{acknowledgements}

\appendix

\section{Implementation of the TM routines} \label{appendix:implementation}

As for RHT, TM is based on a simple set of operations that are repeated identically a
number of times, possibly at every pixel position of the analysed image. Therefore, the
method is well suited for parallel processing and for GPU computations. We have
implemented the programme using the Python language. With the help of the pyOpenCL
library\footnote{\tt https://github.com/pyopencl/pyopencl}, the main computations are
carried out by separate OpenCL \footnote{\tt https://www.khronos.org/opencl/} kernels.
The kernels are compiled pieces of code that are run on a `device', which can be for
example a central processing unit (CPU) or a graphics processing unit (GPU) device. The
code available at {\tt www.interstellarmedium.org/PatternMatching} can be used for 2D
FITS images. The web page includes a basic example of the use of the software.

It is possible to apply TM to maps of different pixelisation and even for all-sky maps.
This could be useful both for scientific analysis and for data quality control. The
method only requires that one is able to read data values at spatial offsets that are
determined by the shape, scale, and orientation of the template. As an example of the
feasibility of analysis of large data sets, we have run TM algorithm on the all-sky
$Planck$ 857\,GHz data. For 50 million pixel position the convolutions (with $F_1$ equal
to 2$\degr$) and the template matching took a total of about 20 seconds.

\section{Alternative templates} \label{appendix:alternative}

\begin{figure}
\includegraphics[width=8cm]{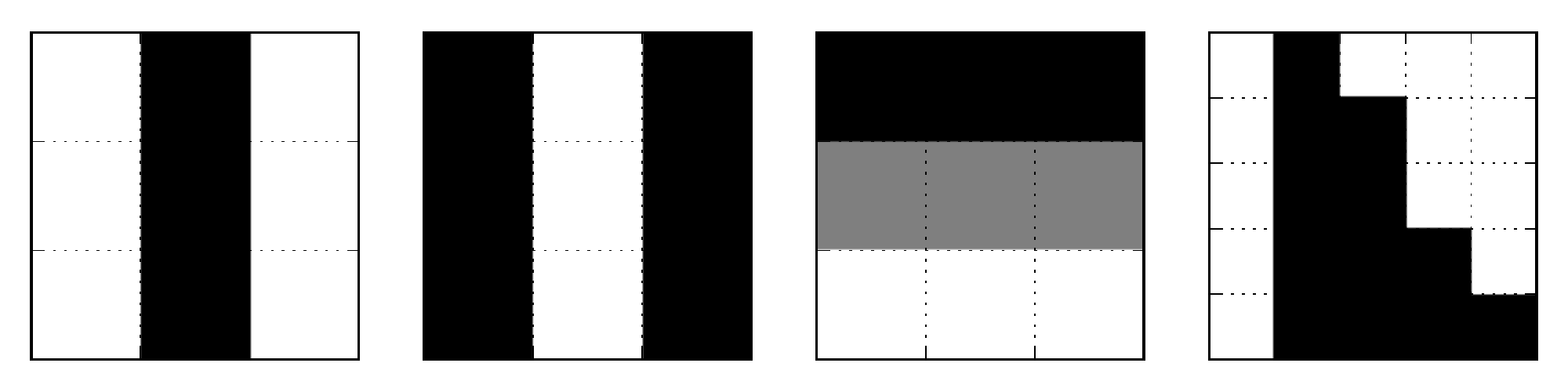}
\caption{
Examples of simple templates. From left to right: filament, valley, gradient, and
wedge. The colours from white to black correspond to values 0, 0.5, and 1. In our
implementation, the final pattern includes an offset that makes the sum of pattern
elements zero.
}
\label{fig:templates}
\end{figure}

\begin{figure}
\includegraphics[width=8.8cm]{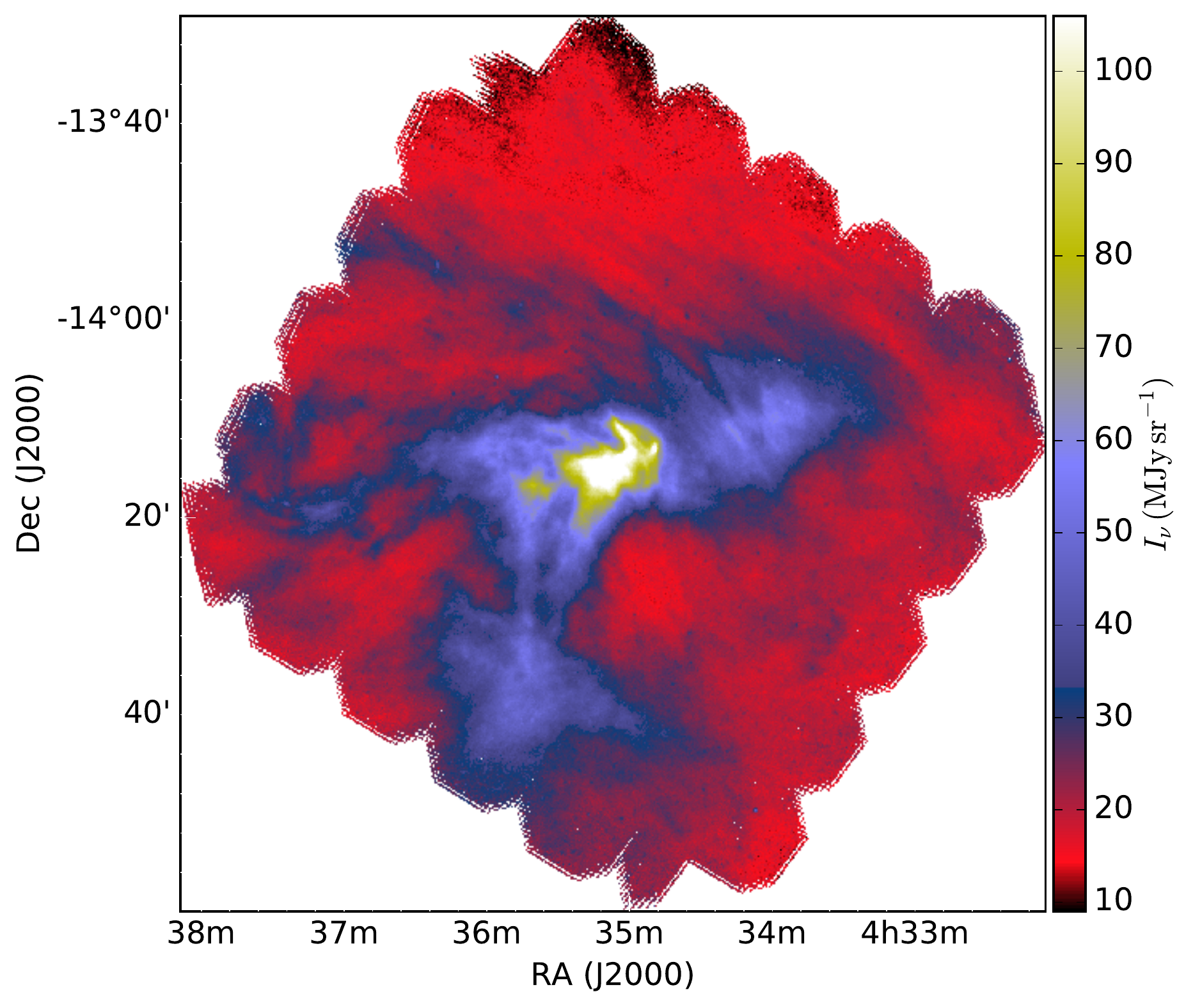}
\caption{
The 250\,$\mu$m surface brightness map of the cloud L1642.
}
\label{fig:L1642}
\end{figure}

Examples in Sect.~\ref{sect:results_RHT} showed that TM can be used to recognise
elongated structures. However, TM is a framework that enables the detection of different
structural features according to the adopted template. The same basic algorithm can also
be used for data with other dimensionality and thus, for example, also for 1D and 3D
applications. As long as the size of the template (in terms of the number of points) is
small, the method remains fast and is thus suitable for mass processing of
data and for exploratory data analysis. We take as examples four 2D templates that we
name filament, valley, gradient, and wedge (see Fig.~\ref{fig:templates}), the
first one being the one already used in this paper. These are applied to the 250\,$\mu$m
surface brightness data of LDN~1642 \citep[field G210.90-36.55, see][]{GCC-III, GCC-IV}
that are shown in Fig.~\ref{fig:L1642}.

\begin{figure*}
\includegraphics[width=18cm]{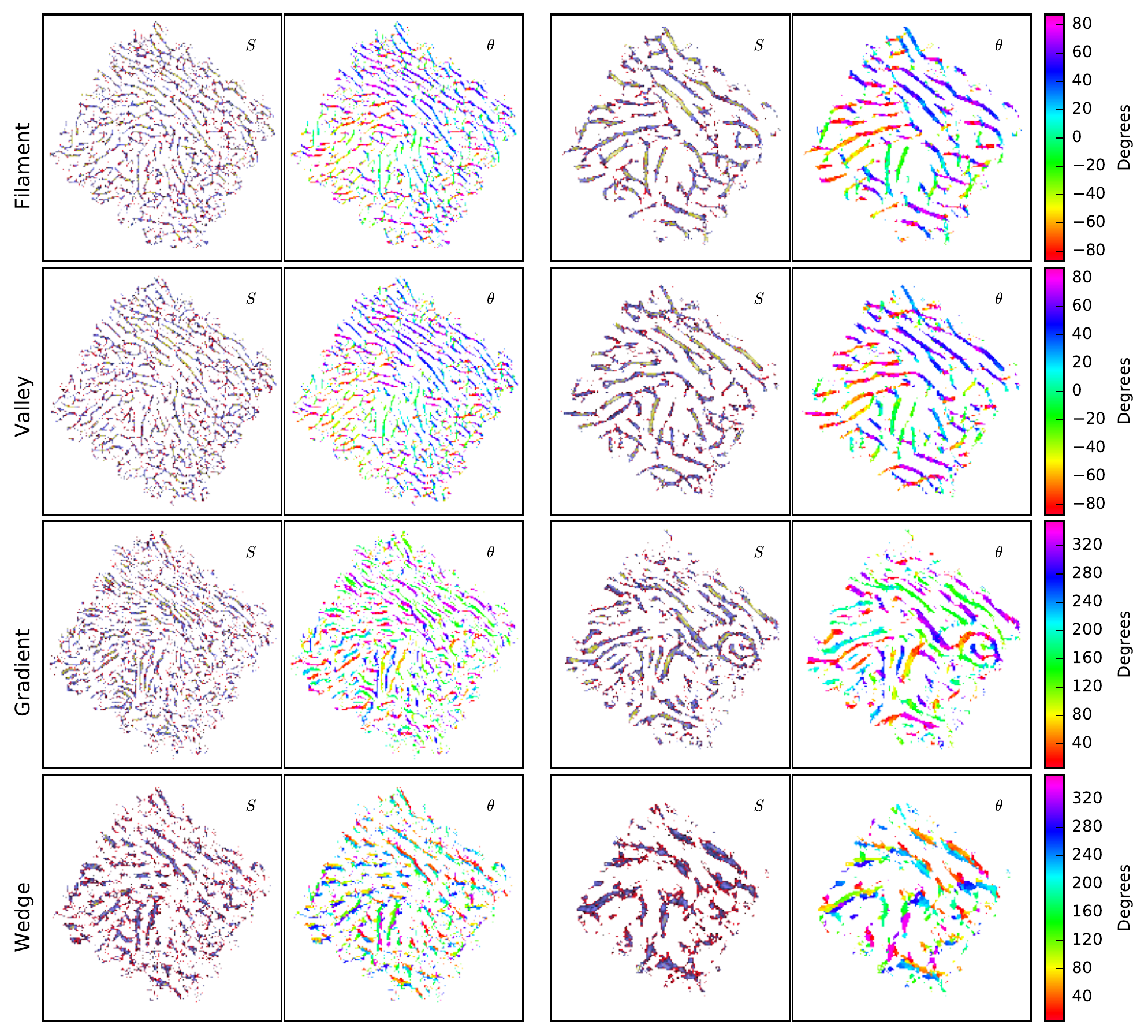}
\caption{
Results of TM using four different templates. The analysed data are the 250\,$\mu$m
surface brightness map of LDN~1642 in Fig.~\ref{fig:L1642}. The four templates (named on
the left side of the figure) are depicted in Fig.~\ref{fig:templates}. Each pair of
frames shows the $S$ and $\theta$ maps of one extraction. On each row, the first two
frames correspond to $F_1=1\arcmin$ and the last two frames to $F_1=2\arcmin$. The
colour scale of $S$ maps is the same as for the position angles, with a different but a
fixed numerical scale. Pixels with $S$ below the 80\% percentile value have been masked.
}
\label{fig:alt_G210_90-36_55_std1_filter1}
\end{figure*}

Figure~\ref{fig:alt_G210_90-36_55_std1_filter1} shows the results at scales of
$F_1=1\arcmin$ and $F_1=2\arcmin$. We use data normalisation (method TMN) which,
together with the spatial filtering, makes the analysis almost equally sensitive to
structures at low and high surface brightness levels. We plot all pixels above the
80\% percentile value of $S$. The fit is done independently at each pixel position. If
the pixels form structures that are long compared to the size of the template, and
especially if the estimated position angles are uniform, the combined detection can be
considered reliable. Especially at the higher $1\arcmin$ resolution, the method with the
filament template is able to highlight many elongated structures that are recognisable
in the input data only by a very careful inspection. One can also recognise coherent
patterns of elongated structures that individually are only barely visible above the
noise. Figure~\ref{fig:alt_G210_90-36_55_std1_filter1} can thus be seen as an aid of
exploratory data analysis although a more quantitative estimation of the reliability of
individual structures is also possible (see Sect.~\ref{sect:reliability}).
The map size is about $90\arcmin \times 90\arcmin$ and the images thus contain a large
number of subregions that are independent at the scale of the templates.

The valley extraction is mainly complementary to the filament and sensitive not only
to real elongated depressions of surface brightness but also to the edges of
filaments. The gradient template is potentially more useful. Without normalisation it
would identify the regions with the largest gradients, providing the absolute values and
the directions of the gradients. Because in
Fig.~\ref{fig:alt_G210_90-36_55_std1_filter1} we are using the method TMN,
the significance $S$ now measures the linearity of the gradients rather than their
magnitude. 

With the last wedge template, we aim to recognise elongated structures whose width
decreases at the scale of the template. It is a more specialised case (and not generic
regarding the opening angle) but might be useful to highlight (for suitable input data)
outflow structures or short striations extruding from a main structure. The result of
wedge template is indeed found to be clearly different from, for example, the filament
template and it does point out some wedge-like structures, especially when applied with
$F_1=2\arcmin$.

\begin{figure*}
\includegraphics[width=18cm]{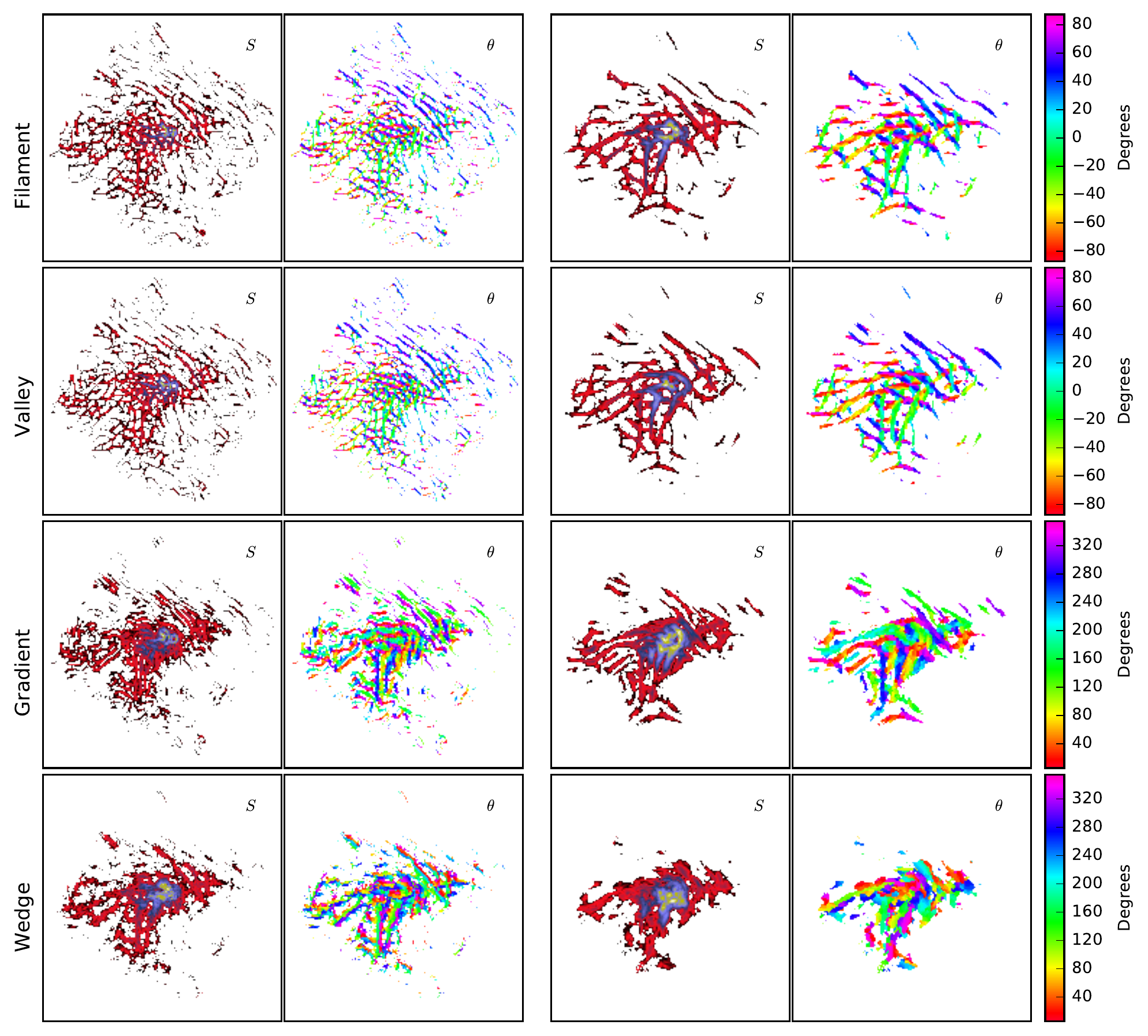}
\caption{
As Fig.~\ref{fig:alt_G210_90-36_55_std1_filter1} but without data normalisation.
%
}
\label{fig:alt_G210_90-36_55_std0_filter1}
\end{figure*}

Figure~\ref{fig:alt_G210_90-36_55_std0_filter1} shows the same results as
Fig.~\ref{fig:alt_G210_90-36_55_std1_filter1} but without data normalisation. In this
case, the method highlights only structures at only high intensity levels.

\bibliography{biblio_with_Planck}

\end{document}